\documentclass[a4paper,11pt]{article}
\usepackage{pos}
\usepackage{hyperref}
\usepackage{graphicx} % for minipage

\title{Study of charm fragmentation with charm meson and baryon azimuthal correlation measurements with ALICE}
\ShortTitle{Study of charm fragmentation with ALICE}

\author*[a,b]{Samuele Cattaruzzi for the ALICE Collaboration}

\affiliation[a]{INFN Sezione di Trieste, Italy}

\affiliation[b]{Dipartimento di Fisica, University of Trieste, Italy}

\emailAdd{samuele.cattaruzzi@cern.ch}

\abstract{Fragmentation functions are typically parametrised exploiting measurements performed in $\mathrm{e^+e^-}$ and $\mathrm{ep}$ collisions, under the assumption of universality across collision systems.
Measurements of charm-hadron yields in proton--proton (pp) collisions at LHC have proved that the hadronisation of heavy quarks differs between hadronic and leptonic collisions. Measurements of charm meson and baryon azimuthal correlations, as well as tagged jets, can provide more stringent constraints on the characteristics of charm hadronisation in hadronic collisions. The ALICE Collaboration has conducted detailed studies on the azimuthal correlation with charged particles, as well as tagged jets measurements of non-strange D mesons and $\Lambda_{\text{c}}^{+}$ baryons in pp collisions at $\sqrt{s} = 13$~TeV. Additionally, ALICE has investigated the azimuthal correlation of $\text{D}_{\text{s}}^{+}$ mesons with charged particles in pp collisions at $\sqrt{s} = 13.6$~TeV. This study provides valuable insights into the possible dependence of charm quark hadronisation on the flavour content of the hadrons produced in the final state. \\
This deeper understanding of charm hadronisation mechanisms, achieved thanks to the contribution of all these studies, can significantly enhance the accuracy of fragmentation function modelling and improve our overall knowledge of quantum chromodynamics in high-energy hadronic collisions.}

\FullConference{42nd International Conference on High Energy Physics (ICHEP2024)\\
18-24 July 2024\\
Prague, Czech Republic\\}

\begin{document}
\maketitle

\section{Introduction}
The production cross sections of heavy-flavour hadrons can be determined using the factorisation approach \cite{hadron production}, which involves the convolution of three components: the parton distribution functions (PDFs) of the incoming protons, the hard-scattering cross section, calculated perturbatively in powers of the strong coupling constant $\alpha_{s}$, and the fragmentation functions (FFs) of heavy quarks into specific hadrons. Fragmentation functions were generally assumed to be universal across different collision systems.

Heavy-flavour baryon-to-meson yield ratios are key observables for probing the hadronisation mechanism, as the contributions from PDFs and hard-scattering cross sections largely cancel out in the ratio. The $\Lambda_{\text{c}}^{+}/\text{D}^{0}$ ratio in proton--proton (pp) collisions at the LHC \cite{LambdaC_over_D0} shows a $p_{\text{T}}$-dependent enhancement compared to $\mathrm{e^+e^-}$ and $\mathrm{ep}$ measurements, questioning the universality of the FFs. While these measurements focus on production yields, studying particles produced in association with charm hadrons can provide deeper insights into the hadronisation mechanism and can refine model predictions. These measurements include azimuthal correlations of charm hadrons with charged primary particles and charm-tagged jets.

In our azimuthal correlation measurement, we probe the relative angle between the directions of charm hadrons and other charged primary particles produced in the collision. At leading order (LO), perturbative QCD predicts that $\text{c}\Bar{\text{c}}$ production results in a back-to-back azimuthal topology at quark level, leading, after the parton shower and the hadronisation, to two peaks in the measured azimuthal distribution: a near-side peak centered at $\Delta \varphi = 0$ and an away-side peak at $\Delta \varphi = \pi$. These peaks represent the azimuthal distribution and particle multiplicity within the jets produced by charm quarks, providing insights into the charm production and hadronisation processes. Next-to-leading order (NLO) production mechanisms can give rise to significantly different correlation patterns \cite{NLO production}\cite{NLO production 2}.

Charm-tagged jet measurements focus on jets containing charm hadrons and analyse the charm-jet production yield as a function of the longitudinal momentum fraction ($z_{||}^{\text{ch}}$) carried by the charm hadron within the jet. These distributions provide information about the hadronisation process of charm quarks.

\section{Non-strange D and $\text{D}_{\text{s}}^{+}$ meson azimuthal correlations}
ALICE measured the azimuthal correlations of non-strange D mesons with charged particles as a function of D-meson transverse momentum ($p_{\text{T}}$) in pp collisions at centre-of-mass energies of $\sqrt{s} = 5.02$ \cite{Ang corr 5 TeV}, 7 \cite{Ang corr 7 TeV}, and 13 \cite{Ang corr 13 TeV} TeV. The azimuthal correlation distributions were fitted with a sum of two generalised Gaussians to model the peaks, along with a constant term, to extract the physical observables such as the peak yields and widths. The results show increasing near-side yields with increasing $p_{\text{T}}$, which is likely due to the larger phase space available for additional charm-fragment production at higher momenta. The near-side peak narrows with increasing $p_{\text{T}}$, indicating that the fragments become more collimated, due to the increased boost of the charm quark. Additionally, measurements at different energies show compatible results, suggesting no significant dependence on collision energy within the studied range. The experimental results were compared with different models. Among these, PYTHIA 8 \cite{PYTHIA8} and POWHEG+PYTHIA 8 \cite{POWHEG+PYTHIA8 3} showed better agreement with the data across a broad $p_{\text{T}}$ range.

\begin{figure}[tb]
\centering
\includegraphics[width=.44\linewidth]{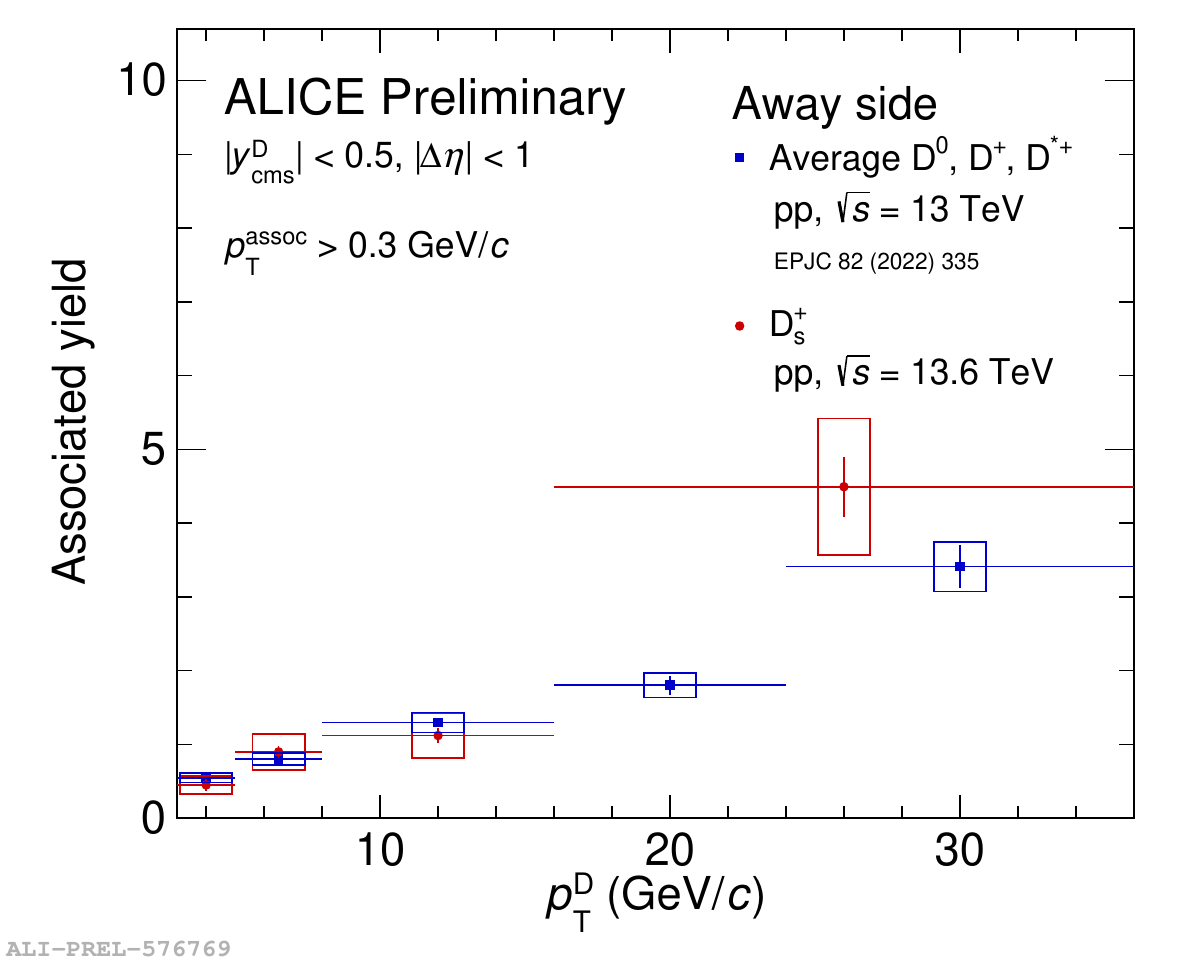} 
\includegraphics[width=.44\linewidth]{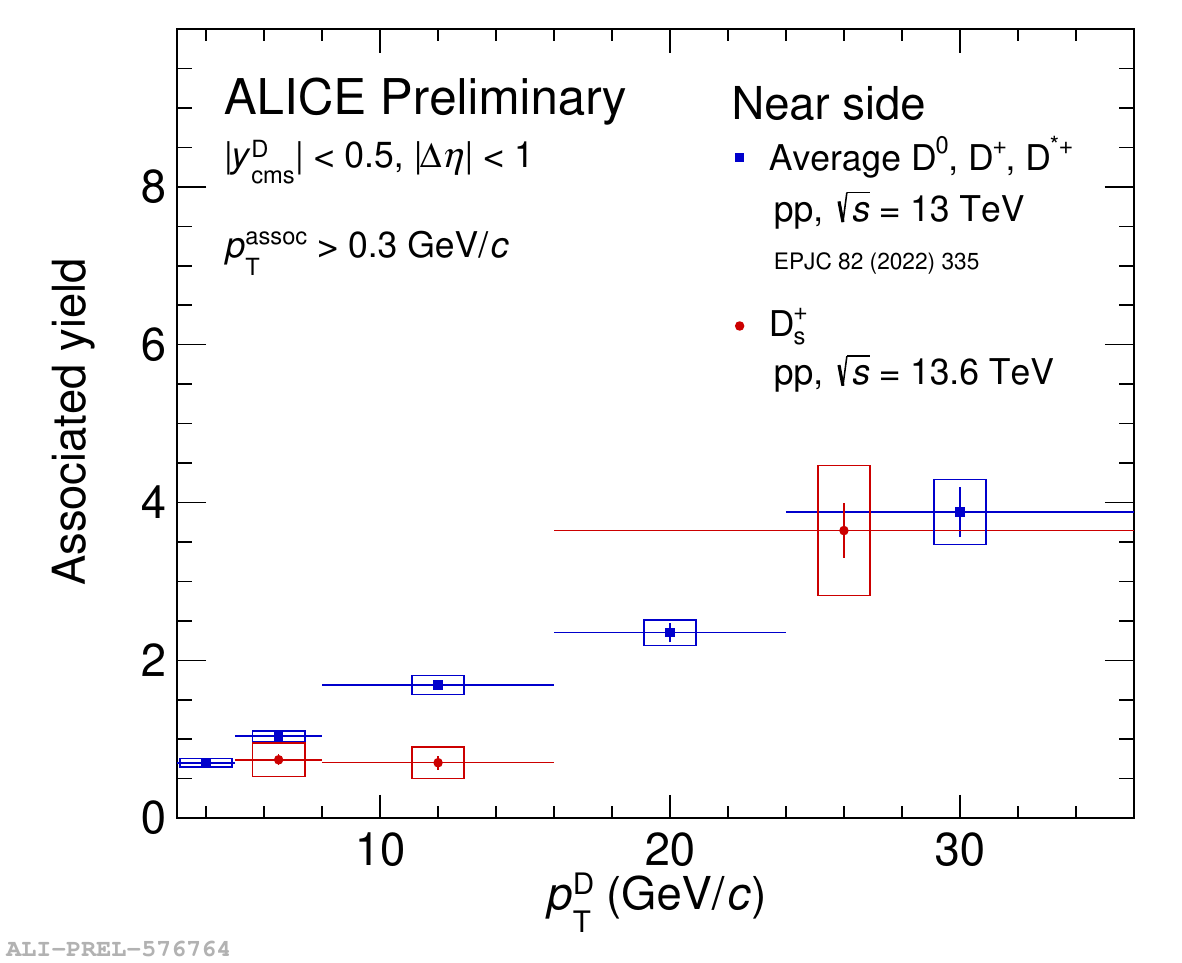} 
\caption{Away-side (left) and near-side (right) yield comparison between the azimuthal correlations of $\text{D}_{\text{s}}^{+}$ and non-strange D mesons with associated charged particles with $p_{\text{T}}^{\text{assoc}}$ > 0.3 GeV/$c$.}
\label{Fig:Away-side-yield-Ds}
\end{figure}

\begin{figure}[tb]
\centering
\includegraphics[width=.44\linewidth]{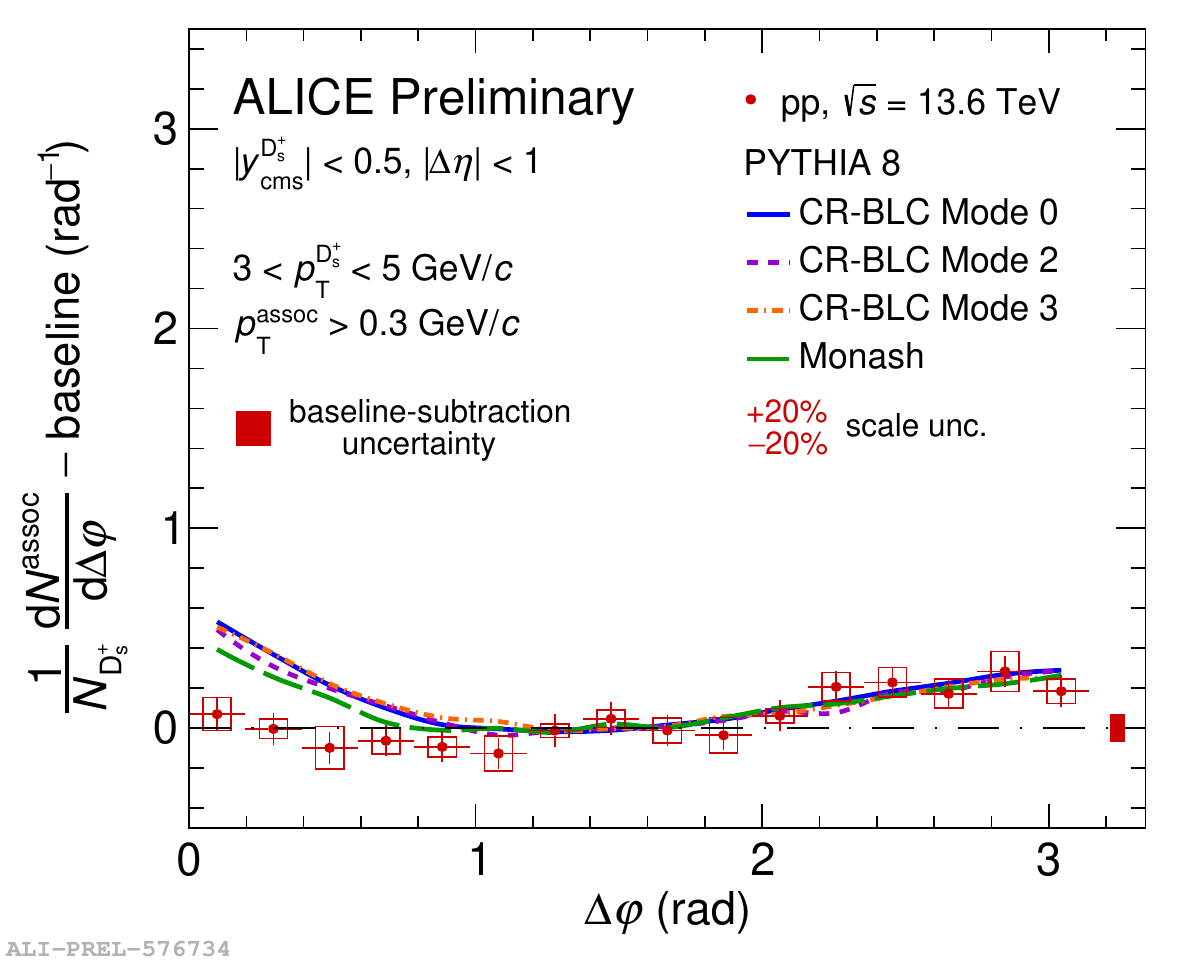} 
\includegraphics[width=.44\linewidth]{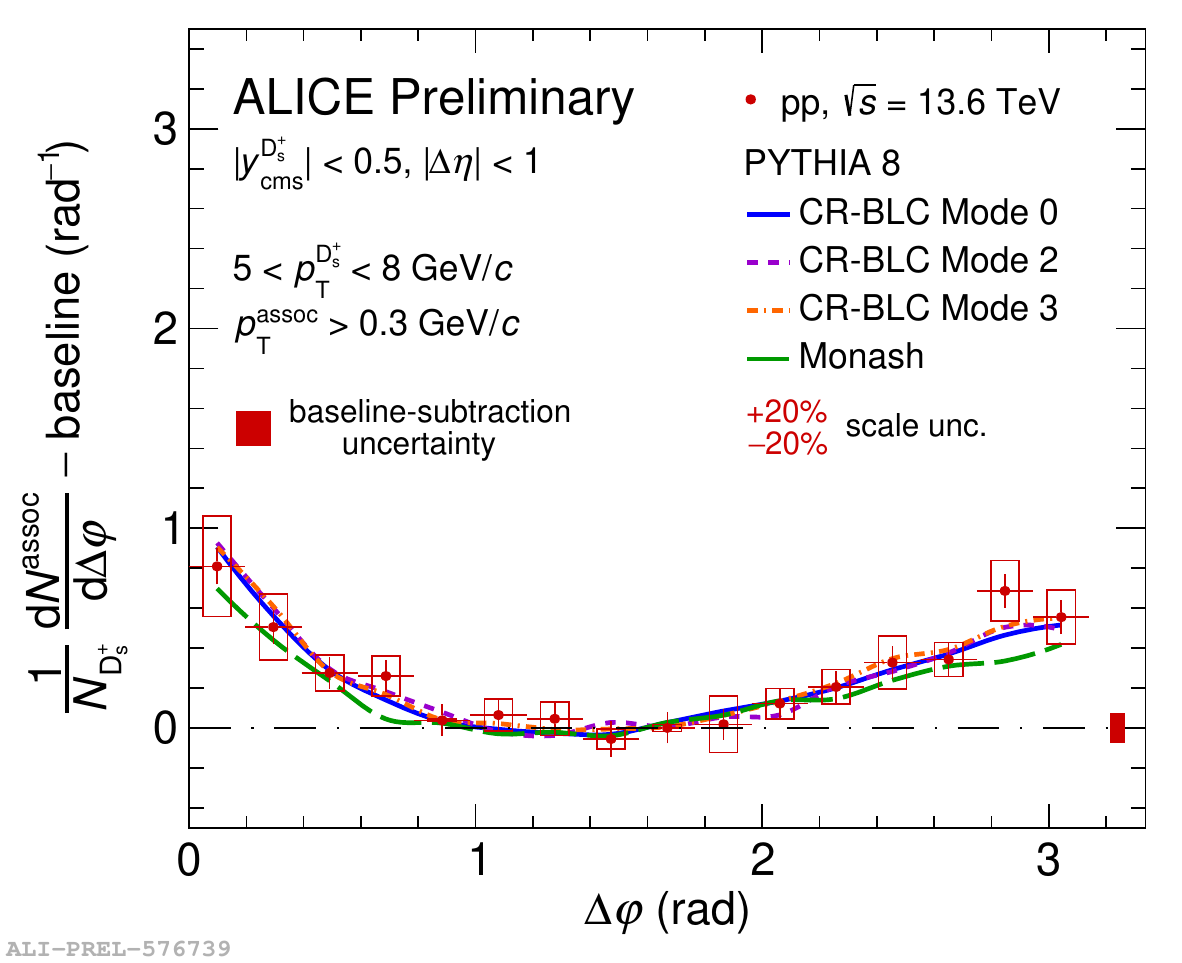} 
\caption{Comparison between the measured $\text{D}_{\text{s}}^{+}$ azimuthal correlation distributions and PYTHIA 8 predictions for 3 < $p_{\text{T}}(\text{D}_{\text{s}}^{+})$ < 5 GeV/$c$ (left) and 5 < $p_{\text{T}}(\text{D}_{\text{s}}^{+})$ < 8~GeV/$c$ (right), in both cases for $p_{\text{T}}^{\text{assoc}}$ > 0.3 GeV/$c$.}
\label{Fig:Ang-corr-model-comp}
\end{figure}

ALICE extended this study to $\text{D}_{\text{s}}^{+}$ mesons using a sample of pp collisions at $\sqrt{s} = 13.6$ TeV collected during LHC Run 3. Figure \ref{Fig:Away-side-yield-Ds} presents a comparison of the away- and near-side yields from azimuthal correlations of non-strange D and $\text{D}_{\text{s}}^{+}$ mesons with charged particles.  From the comparison, a good agreement over the full measured $p_{\text{T}}$ range is found for the away-side peak yields. However, up to 4$\sigma$ difference is observed for the near-side peak, where the $\text{D}_{\text{s}}^{+}$ meson yields for $p_{\text{T}}$~<~16 GeV/$c$ are lower than those for non-strange D mesons, suggesting a possible difference in hadronisation processes due to the different strangeness content of the two species. In Figure \ref{Fig:Ang-corr-model-comp} the comparison between the measured $\text{D}_{\text{s}}^{+}$ azimuthal correlations and calculations from the PYTHIA 8 Monash \cite{PYTHIA Monash} and CR-BLC \cite{PYTHIA CR BLC} tunes is reported. Model comparisons show good agreement for 5~<~$p_{\text{T}}$~<~8~GeV/$c$, while significant discrepancies are evident for $p_{\text{T}}$~<~5~GeV/$c$ in the near-side peak, suggesting potential areas for further improvements in theoretical modelling.

%\begin{figure}[tb]
%   \begin{minipage}{0.48\textwidth}
%     \centering
%     \includegraphics[width=.98\linewidth]{Images/dsobs_compwnonstrangd_asyield.pdf}
%     \caption{Away-side yield comparison between $\text{D}_{\text{s}}^{+}$ and non-strange D mesons azimuthal correlations for $p_{\text{T}}^{\text{assoc}}$ > 0.3 GeV/$c$.}\label{Fig:Away-side-yield-Ds}
%   \end{minipage}\hfill
%   \begin{minipage}{0.48\textwidth}
%     \centering
%     \includegraphics[width=.98\linewidth]{Images/dsobs_compwnonstrangd_nsyield.pdf}
%     \caption{Comparison between $\text{D}_{\text{s}}^{+}$ baryon and non-strange D mesons azimuthal correlation distributions measured in the transverse momentum interval 3 < $p_{\text{T}}(\text{D})$ < 5 GeV/$c$  and $p_{\text{T}}^{\text{assoc}}$ > 0.3 GeV/$c$.}\label{Fig:Ds-ang-corr-comparison}
%   \end{minipage}
%\end{figure}

\section{$\text{D}^{0}$ and $\text{D}_{\text{s}}^{+}$  meson tagged jets measurements}
\begin{figure}[tb]
\centering
\includegraphics[width=.44\linewidth]{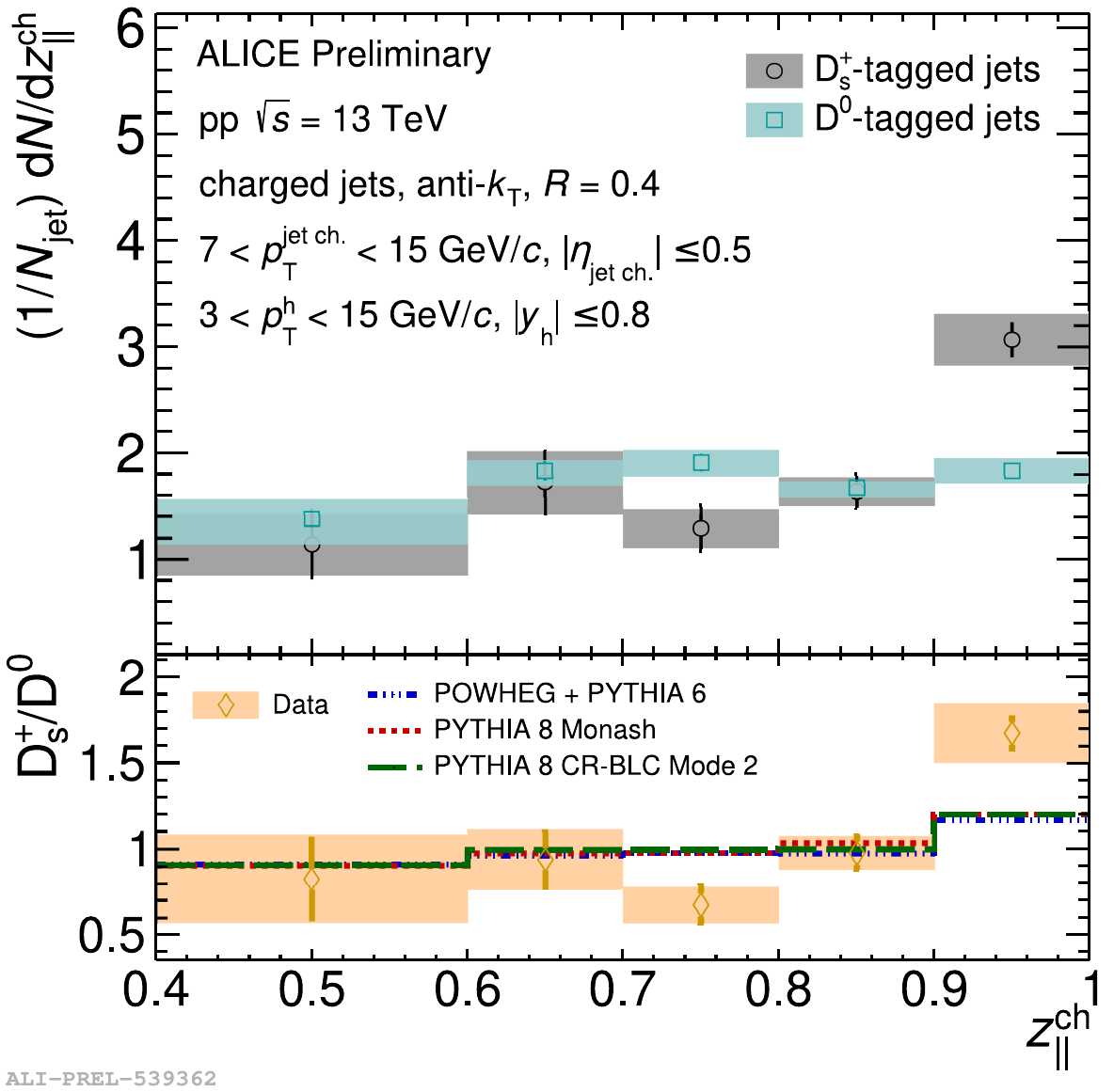}
\caption{Top panel: Comparison of the yields of $\text{D}_{\text{s}}^{+}$ and $\text{D}^{0}$-tagged jets as a function of the longitudinal jet momentum fraction for 3 < $p_{\text{T}}(\text{D}_{\text{s}}^{+}, \text{D}^{0})$ < 15 GeV/$c$ and 7 < $p_{\text{T}}^{\text{jet ch.}}$ < 15 GeV/$c$. Bottom panel: Comparison of the $z_{||}^{\text{ch}}$ distribution ratio for $\text{D}_{\text{s}}^{+}$/$\text{D}^{0}$-tagged jets with different model predictions.}
\label{Fig:Ds-tagged-jets}
\end{figure}

Further insights into charm-quark hadronisation can be obtained from the study of charm-tagged jet measurements. ALICE reconstructed $\text{D}^{0}$ and $\text{D}_{\text{s}}^{+}$-tagged jets in pp collisions at $\sqrt{s} = 13$~TeV, measuring the longitudinal momentum fraction ($z_{||}^{\text{ch}}$) carried by the charm hadron within the jet. These studies provide valuable information on how charm quarks fragment and hadronise into different final states. When comparing the results of $\text{D}^{0}$- and $\text{D}_{\text{s}}^{+}$-tagged jets, a higher yield is observed for $\text{D}_{\text{s}}^{+}$-tagged jets compared to $\text{D}^{0}$-tagged jets at values of $z_{||}^{\text{ch}}$ closer to unity, as shown in Fig. \ref{Fig:Ds-tagged-jets}. This observation could suggest a harder hadronisation of charm quarks into $\text{D}_{\text{s}}^{+}$ mesons than into $\text{D}^{0}$ mesons in the studied kinematic range, and in general points to differences in the fragmentation process or hadronisation dynamics between the two mesons. Additionally, PYTHIA 8 predictions with Monash and CR-BLC tunes reproduce the measured $\text{D}_{\text{s}}^{+}/\text{D}^{0}$ ratio within uncertainties, except for the last $z_{||}^{\text{ch}}$ interval, where a discrepancy arises.

\section{$\Lambda_{\text{c}}^{+}$ azimuthal correlations and tagged-jet measurements}
\begin{figure}[tb]
\centering
\includegraphics[width=.72\linewidth]{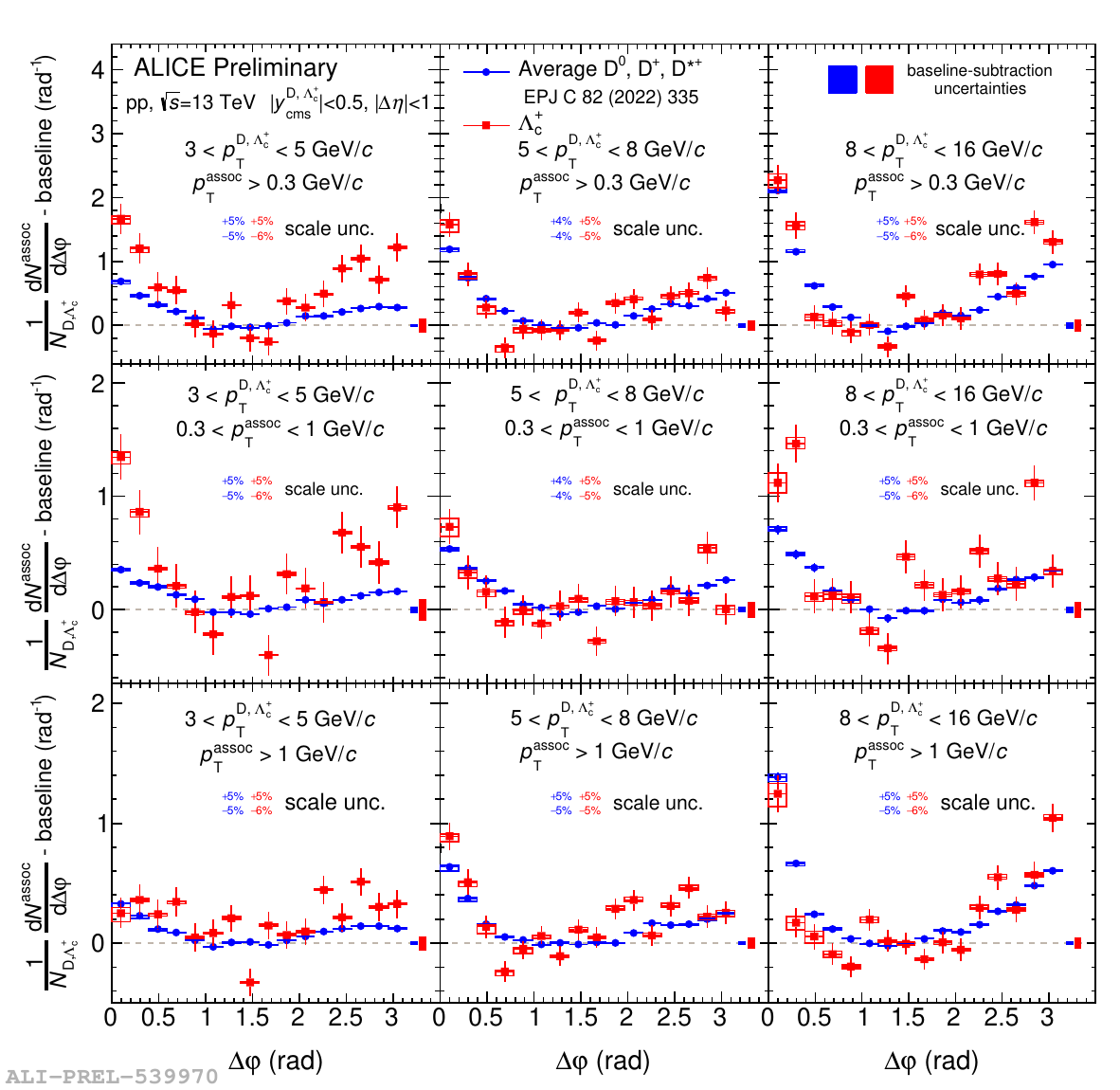}
\caption{Comparison between $\Lambda_{\text{c}}^{+}$-triggered and non-strange D-meson-triggered azimuthal correlation distributions with charged particles measured in the transverse momentum intervals 3~<~$p_{\text{T}}(\text{D}, \Lambda_{\text{c}}^{+})$~<~5~GeV/$c$, 5~<~$p_{\text{T}}(\text{D}, \Lambda_{\text{c}}^{+})$~<~8 GeV/$c$, 8~<~$p_{\text{T}}(\text{D}, \Lambda_{\text{c}}^{+})$~<~16~GeV/$c$ (from left to right) and $p_{\text{T}}^{\text{assoc}}$~>~0.3~GeV/$c$, 0.3~<~$p_{\text{T}}^{\text{assoc}}$~<~1~GeV/$c$, $p_{\text{T}}^{\text{assoc}}$~>~1~GeV/$c$ (top, central and bottom panels, respectively).}
\label{Fig:LambdaC}
\end{figure}

ALICE measured the azimuthal correlations between $\Lambda_{\text{c}}^{+}$ baryons and charged particles, as well as $\Lambda_{\text{c}}^{+}$-tagged jets, in pp collisions at $\sqrt{s} = 13$ TeV. These measurements aim to provide insights into charm-quark fragmentation into $\Lambda_{\text{c}}^{+}$ baryons. In Fig. \ref{Fig:LambdaC} the comparison between $\Lambda_{\text{c}}^{+}$-triggered and non-strange D-meson-triggered azimuthal correlation distributions with charged particles measured in three $p_{\text{T}}(\text{D}, \Lambda_{\text{c}}^{+})$ intervals from 3 to 16 GeV/$c$ for $p_{\text{T}}^{\text{assoc}}$ > 0.3 GeV/$c$, 0.3 < $p_{\text{T}}^{\text{assoc}}$ < 1 GeV/$c$, $p_{\text{T}}^{\text{assoc}}$ > 1 GeV/$c$ are reported. By comparing the azimuthal correlation distributions, it is observed that for $p_{\text{T}}(\text{D}, \Lambda_{\text{c}}^{+})$ > 5~GeV/$c$, the azimuthal correlations between the two hadron species show a good agreement, indicating similar charm fragmentation behaviour at high $p_{\text{T}}$. However, at lower $p_{\text{T}}$ values, a tendency of an enhancement of both the near- and away-side peaks for $\Lambda_{\text{c}}^{+}$ baryons compared to D mesons is observed. This could suggest a potentially softer charm fragmentation, where greater momentum is distributed among a larger number of particles when forming a $\Lambda_{\text{c}}^{+}$ baryon.

Additionally, $\Lambda_{\text{c}}^{+}$-tagged jet measurements show a hint of a softer jet-momentum fraction ($z_{||}^{\text{ch}}$) than $\text{D}^{0}$-tagged jets \cite{LambdaC tagged jets}. This result indicates that, within the studied kinematic range, charm quarks are more likely to fragment into $\Lambda_{\text{c}}^{+}$ baryons than $\text{D}^{0}$ mesons when a moderate fraction of the jet momentum is carried by the hadron. The measured $\Lambda_{\text{c}}^{+}/\text{D}^{0}$ ratio is compared with the predictions from PYTHIA 8 CR-BLC and Monash tunes, with PYTHIA 8 CR-BLC showing better agreement with the data than the Monash tune.

\section{Conclusions}
The ALICE Collaboration carried out a detailed study of charm-quark fragmentation using charm meson and baryon azimuthal correlations as well as charm-hadron tagged-jet measurements. For non-strange D mesons, the results show increasing near-side yields and narrowing peak widths with rising transverse momentum ($p_{\text{T}}$), indicating more collimated jets at higher $p_{\text{T}}$. These findings are consistent across different collision energies ($\sqrt{s} = 5.02$, 7, and 13 TeV) and are reproduced by PYTHIA 8 and POWHEG+PYTHIA 8 predictions.

For $\text{D}_{\text{s}}^{+}$ mesons, the away-side peak shape is compatible with non-strange D mesons. However, lower near-side yields at low $p_{\text{T}}$ could hint to a different hadronisation pattern due to strangeness. Additionally, $\text{D}_{\text{s}}^{+}$-tagged jet measurements suggest a harder fragmentation of the $\text{D}_{\text{s}}^{+}$ compared to the $\text{D}^{0}$, with higher yields at longitudinal momentum fractions closer to unity.

For $\Lambda_{\text{c}}^{+}$ baryon azimuthal correlations, a possible enhancement of near-side and away-side peaks at low $p_{\text{T}}$ and higher yields at intermediate momentum fractions is observed. This could indicate a softer fragmentation pattern compared to $\text{D}^{0}$ mesons, with the parton momentum being spread over more particles.

\end{document}